# Silicon-organic hybrid thermo-optic switch based on a slot waveguide directional coupler

Li-Yuan Chiang,[1,*] Chun-Ta Wang,[2] Steve Pappert,[1] and Paul K. L. Yu[1]

[1]*Department of Electrical and Computer Engineering, University of California San Diego, La Jolla, CA 92122, USA*
[2]*Department of Photonics, National Sun Yat Sen University, Kaohsiung 80424, Taiwan*
*\*Corresponding author: l1chiang@eng.ucsd.edu*

We propose and demonstrate a passively biased 2 × 2 thermo-optic switch with high power efficiency and fast response time. The device benefits from the highly concentrated optical field of a slot waveguide mode and the strong thermo-optic effect of a nematic liquid crystal (NLC) cladding. The NLC fills the nano-slot region and is aligned by the subwavelength grating inside. The measured power consumption and thermal time constant are 0.58 mW and 11.8 µs, respectively, corresponding to a figure-of-merit of 6.8 mW·µs. The proposed silicon-organic hybrid device provides a new solution to design thermo-optic actuators having low power consumption and fast operation speed. © 2022 Optica Publishing Group.

Silicon photonics is a promising field for various applications, including optical communications [1-2], light detection and ranging (LiDAR) [3-4], artificial intelligence (AI) [5], and programmable photonic integrated circuits (PIC) [6]. Thermo-optic (TO) devices, such as switches and phase shifters, are commonly employed in silicon PICs due to low loss, simple fabrication procedures, and the feasibility of large-scale integration [2]. However, the power consumption of TO devices is a critical challenge for large-scale applications. The maximum number of on-chip TO devices is limited by the chip's power budget and the power consumption of the individual TO devices [7-8]. Silicon photonic TO devices consist of heaters and silicon waveguides based on silicon-on-insulator (SOI) wafers. Although silicon is generally considered to have a large thermo-optic coefficient (TOC) ($1.86 \times 10^{-4}$/°C at 1550 nm wavelength) [9], the thermal spreading from the heaters in the unwanted directions inevitably causes excess power dissipation [10]. Several techniques have been demonstrated to design power efficient TO devices. The heaters can be placed close to the target waveguide with a potential cost of higher loss [11]. Introducing an air trench/undercut is another way to reduce the excess power consumption, but the speed is sacrificed [12]. Using suspended phase arms and a heater-on-slab structure, a low-power and high-speed TO switch has been demonstrated [13]. However, the mechanical robustness of the suspended structure emerges as a challenge. Resonant TO designs (rings [14] and nanobeam cavities [15]) are successful in demonstrating low power consumption and a small footprint. However, the narrow optical bandwidth and poor fabrication tolerance of resonant TO designs limit their applications [11]. As a result, interferometer-based TO designs are preferred in most cases despite consuming more power. Other waveguide materials, such as silicon oxycarbide (SiOC, $2.5 \times 10^{-4}$/°C at 1550 nm wavelength) [16], were proposed for low-power TO device design due to their superior TOC, although the fabrication and integration with SOI wafers present additional challenges.

Silicon-organic hybrid (SOH) devices have gained significant advancement in the last decade, particularly in electro-optic (EO) devices [17-22]. Highly efficient modulation is realized through integrating organic materials, such as EO polymers [17-20] or nematic liquid crystals (NLC) [21-23], with strong EO responses into silicon slot waveguides. Inspired by the fruitful results in SOH EO devices, we previously reported a highly sensitive SOH temperature sensor combining NLC materials and silicon slot waveguides [24]. Although an external electric field is required to actively align the NLC in the slot, the work showed the strong potential of the SOH technique with NLC for TO applications.

In this letter, we propose a power efficient TO switch enabled with a silicon slot waveguide and NLC cladding. The silicon structures passively aligned the NLC in the slot region for the TE propagating light to probe the large TO coefficient of NLC on the extraordinary axis ($dn_e/dT$, on the order of $-2 \times 10^{-3}$/°C at 1550 nm wavelength [25]). Compared with the active alignment method used in [24], the passive one simplifies the device's fabrication process and electrical control. The proposed device possesses low power consumption without compromising the operating speed.

As schematically shown in Fig. 1(a), the TO switch relies on an SOI slot waveguide directional coupler with NLC as the infilling material. The slot waveguide region has a strip-loaded structure with slabs laterally connected to the outside of the silicon coupled arms [26]. The two silicon coupled arms are connected with a silicon subwavelength grating (SWG), forming rectangular nano-grooves inside the slot, Fig. 1(a), Inset(1). The NLC material fills the slot region and the nano-grooves inside. TE-polarized light enters the device from one of the input strip waveguides and launches two supermodes, TE0 and TE1 [24]. The simulated modal distributions of the two supermodes are shown in Fig. 1(b)-(c). The TE0 mode has a maximum field in the slot, whereas the TE1 mode has a minimum field. The distinct modal distributions result in the change of the two-mode interference when the slot index is TO modulated.

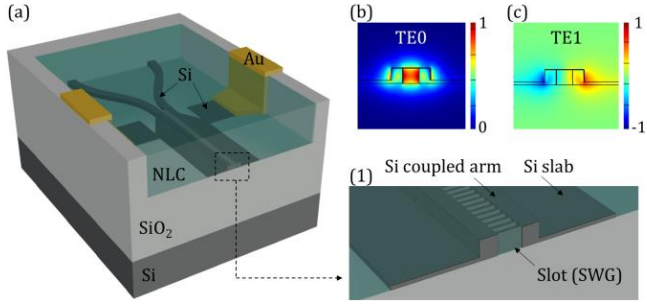

Fig. 1. (a) Schematic of the SOH directional-coupler TO switch. Inset (1): Zoom-in view of the slot waveguide cross section. Simulated modal distributions of the (b) symmetric TE0 mode and the (c) anti-symmetric TE1 mode.

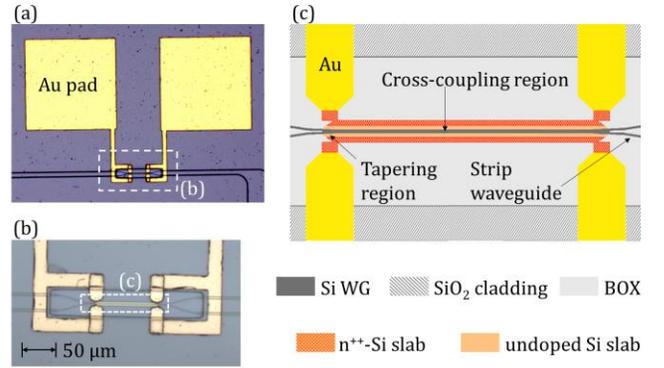

Fig. 2. Top-view (a)-(b) optical microscope images and (c) schematic of the TO switch before applying NLC.

The heater configuration of the TO switch is illustrated in Fig. 2. Two square metal pads were formed for probing and applying electrical current, Fig. 2(a). The metal lines mainly lie on top of the SiO2 cladding. Each metal pad was connected to two silicon square slabs across and covering a 2 μm step height between the SiO2 cladding and the BOX layer. Although there are four crossing parts between the metal lines and silicon waveguides, Fig. 2(b), the optical loss induced is negligible due to the separation in the vertical direction. Fig. 2(c) shows two integrated n++ silicon heaters formed by partially doping the silicon slabs along the slot waveguide direction. The switch is thermally activated via resistive heating at the heaters. The generated thermal energy is transferred to the slot, the sensing hotspot, and induces the TO effect. The figure of merit (FOM) of TO switches is commonly defined as the power-time product [27],

$$\text{FOM} = P_\pi \cdot \tau , \quad (1)$$

where $P_\pi$ is the power consumption required to achieve a π-phase shift, and $\tau$ is the limiting thermal time constant for the switching.

NLC has rod-shaped molecules, strong birefringence, and contrasting TO responses between the extraordinary and ordinary axes. In the nematic phase, the molecules align almost parallel to each other because of inter-molecule forces. 5CB (made by Merck) is the NLC material used for the demonstration in this letter due to its commercial availability, large $dn_e/dT$, and thermal stability [28]. As plotted in Fig. 3 [25], the extraordinary refractive index $n_e$ is highly dependent on temperature, whereas the ordinary refractive index $n_o$ is almost flat in the room-temperature range. <n> is the average refractive index. Since the TO response of the NLC media is determined by the orientation of the molecules, proper alignment is critical to NLC devices. For conventional slot waveguides without an external electric field, the NLC molecules in the slot region align with the slot waveguide due to the sidewall anchoring [21]. However, in this case, both TE and TM modes in the slot waveguide will probe the $n_o$ and small $dn_o/dT$ of 5CB. For the TE modes to sense the $n_e$ and large $dn_e/dT$, the NLC in the slot region needs to be aligned perpendicular to the slot waveguide.

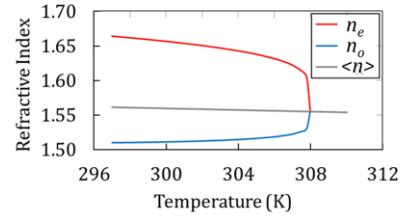

Fig. 3. Temperature-dependent refractive indices of 5CB at 1550 nm wavelength.

With the sidewall anchoring mechanism, it was reported that the NLC orientation could be effectively controlled by nano-groove structures [29-30]. In the proposed design, a nano-groove array is constructed by silicon SWG and the coupled waveguide arms in the slot region. The nano-groove array is perpendicular to the slot waveguide to align the NLC horizontally. To minimize the loss caused by diffraction and scattering, the silicon grating is designed to be in the subwavelength regime with a small and constant grating pitch $\Lambda$ (=105nm), fulfilling the following condition

$$\Lambda \ll \lambda_{eff} = \frac{\lambda}{2n_{eff}} , \quad (2)$$

where $\lambda_{eff}$ is the effective wavelength of the waveguide mode, and $n_{eff}$ is the effective index. Such SWG is equivalent to an effective medium and supports low-loss propagation [31]. Besides realizing passive NLC alignment, the silicon SWG assists in the heat transfer in the slot region, as silicon has a much higher thermal conductivity ($k_{Si}$ = 148 W/m·K [32]) compared to that of 5CB ($k_{5CB}$ ~ 0.2 W/m·K [33]).

The fabrication process is illustrated in Fig. 4. An SOI wafer with a 250 nm thick silicon device layer and a 3 μm thick SiO2 buried oxide (BOX) layer was used. The silicon strip-loaded waveguide and the SWG were formed simultaneously using two steps of electron-beam lithography (EBL) followed by reactive ion etching (RIE) with SF6:CHF3 plasma. The outer parts of the slabs were heavily doped with phosphorus to become n++ silicon heaters. Each heater is connected with square slabs at both ends for metal contacts. A 2 μm SiO2 layer is deposited by plasma-enhanced chemical vapor deposition (PECVD) on top of the whole chip. The slot region was opened through etching. Gold contact pads were formed by sputtering and lift-off. NLC was applied to the slot region and filled the nano-groove array through its liquid nature as the final step.

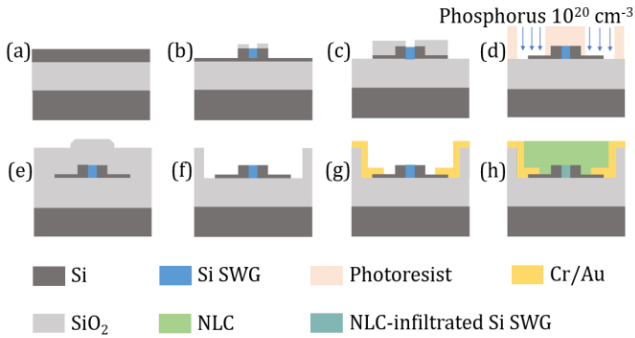

Fig. 4. Schematic representation of the fabrication flow.

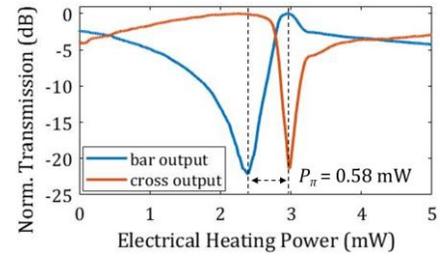

Fig. 6. Measured optical transmission results as a function of electrical heating power applied.

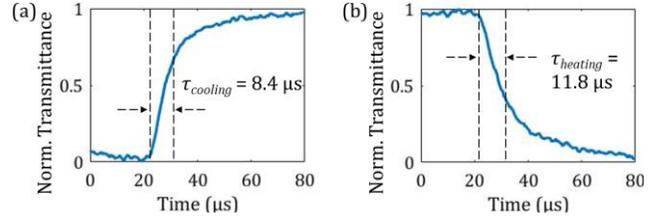

Fig. 7. Measured transient responses of the TO switch for (a) switching ON (cooling) and (b) switching OFF (heating).

Top-view SEM images of the fabricated TO switch before NLC infiltration are shown in Fig. 5. The silicon square slab regions (5 μm on each side) are designed for reliable photolithography alignment during metal contact formation, Fig. 5(a). The device length $L$ is 80 μm, plus additional 5 μm long tapering regions for adiabatic optical mode transition at the directional coupler's input and output sides. The horizontal nano-grooves are nearly rectangular, with an aspect ratio of 4 (~75 × 300 nm$^2$), Fig. 5(b). The width of the coupled waveguide arms is 210 nm. The $\Lambda$ and the silicon filling factor of the SWG are 105 nm and 0.29, respectively.

The optical transmission was measured from both output ports (cross and bar) in the 2×2 switch. The optical measurement setup is illustrated and described in Supplement 1. The normalized transmission results for different total heating power are plotted in Fig. 6. The measured average electrical power consumption $P_\pi$ is ~0.58 mW for both output ports with extinction ratios of ~21 dB. The measured total resistance is 10.7 kΩ, and the corresponding half-wave voltage $V_\pi$ is 0.57 V, the voltage required to achieve a π-phase shift. The device insertion loss is ~1.3 dB. The 5CB became isotropic with nearly flat transmission curves as the power was increased beyond ~3.2 mW, corresponding to the temperature range above the clearing point (308 K) of 5CB shown in Fig. 3.

The transient responses are shown in Fig. 7, with the results measured using a square-wave driving signal at 5 kHz. The cooling (rise) time constant $\tau_{cooling}$ and heating (fall) time constant $\tau_{heating}$ measured are 8.4 μs and 11.8 μs, respectively. It is noteworthy that tunable NLC devices relying on controlling the reorientations of NLC molecules usually have response time at the millisecond level [21-22]. In contrast, the speed of the proposed Si-NLC hybrid TO switch depends on the thermal properties of both silicon and the NLC material incorporated, leading to a response time at the microsecond level. The slower $\tau_{heating}$ compared to the $\tau_{cooling}$ is as expected due to the nonlinear TO response of the NLC. As shown in the transfer curve in Fig. 6, the transmission slopes became sharper when the power was increased up to ~3 mW. Therefore, the heating process requires a larger temperature change and time to reach the 1/$e$ signal level defined by $\tau$.

The performance factors of the proposed switch and state-of-the-art TO switches are listed in Table. 1. The calculated FOM for the proposed directional-coupler TO switch is 6.8 nW·s, which is seven times better than the conventional and optimized TO switch with ridge waveguide phase shifters. The switch enabled by laterally supported suspended phase arms and efficient electrodes was reported to have a better FOM of 5.0. However, the mechanical robustness of the suspended structures remains a concern. Our proposed technique has reached a comparable FOM without suspended structures and can be further improved in many ways. The NLC material used in this work is commercially available. Further material investigation can lead to NLCs with a larger TO effect. The silicon waveguide and heater geometry design can be optimized for better power efficiency and faster heat transfer. Combining the spiral waveguide design and the proposed SOH design is another potential direction for further FOM enhancement.

It can be seen in Table 1 that the Mach-Zehnder interferometer (MZI) is the most common structure design for other reported TO switches. However, the directional-coupler switch configuration is competitive in terms of compact and low-loss design when the index tuning effect is sensitive, as analyzed in [36]. For the MZI switches, the additional footprint and insertion loss induced by the splitter/combiner pair and routing waveguides are inevitable but may not be included in the numbers reported.

In conclusion, a Si-NLC directional-coupler TO switch was proposed and demonstrated with a low power consumption of 0.58 mW, facilitated by the strong TO response of the NLC material incorporated in the slot waveguide. The NLC was passively aligned in the slot region by SWG without extra fabrication steps or external

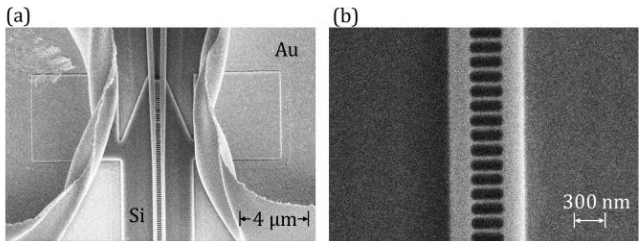

Fig. 5. (a)-(b) SEM images of the TO switch before applying NLC.

**Table 1. Characteristics of the State-of-the-Art Interferometric Thermo-Optic Switches**

| Device design | Interferometer | $L$ (μm) | $P_\pi$ (mW) | $\tau$ (μs) | FOM (mW·μs) | $V_\pi$ (V) | Heater |
|---|---|---|---|---|---|---|---|
| Conventional and optimized [11] | MZI | 320 | 22.8 | 2.2 | 50.2 | 4.8 | n$^{++}$-Si |
| Adiabatic bend [27] | MZI | ~10 | 12.7 | 2.4 | 30.5 | 12 | n$^{++}$-Si |
| Carrier injection w/ heat sink [34] | MZI | 16 | 22.6 | 0.5 | 11.3 | 17.6 | p-Si |
| Spiral waveguide [35] | MZI | 67 | 3.0 | 5.0 | 15.0 | 1.4 | Ti |
| Multi-pass waveguide [8] | MZI | 870 | 1.7 | 6.6 | 11.2 | - | W |
| Suspended [12] | MZI | 1000 | 0.5 | 65.5 | 32.1 | 0.9 | TiN |
| Suspended and laterally supported [13] | MZI | 100 | 1.07 | 4.7 | 5.0 | 0.2 | Cr/Au |
| Si-NLC(5CB) (This work) | Directional coupler | 90* | 0.58 | 11.8 | 6.8 | 0.57 | n$^{++}$-Si |

*It represents the total length of the directional-coupler switch, but the $L$ only represents the phase shifter length for the Mach-Zehnder Interferometer (MZI) switches.

alignment fields. An FOM of 6.8 mW·μs was achieved, which is seven times better than the optimized TO switch using a conventional ridge phase shifter with n$^{++}$ silicon heaters. The proposed SOH technique paves a new way for developing power efficient TO devices with potentially single-digit microsecond responses.

**Funding.** Office of Naval Research (contract N00014-18-1-2027).

**Disclosures.** The authors declare no conflicts of interest.

**Data availability.** Data underlying the results presented in this paper are not publicly available at this time but may be obtained from the authors upon reasonable request.

**Supplemental document.** See Supplement 1 for supporting content.

## References

1. Q. Cheng, M. Bahadori, M. Glick, S. Rumley, and K. Bergman, Optica **5**, 1354-1370 (2018).
2. B. G. Lee and N. Dupuis, J. Light. Technol. **37**, 6-20 (2019).
3. C.-P. Hsu, B. Li, B. Solano-Rivas, A. R. Gohil, P. H. Chan, A. D. Moore, and V. Donzella, IEEE J. Sel. Top. Quantum Electron. **27**(1), 1-16 (2021).
4. C. Li, X. Cao, K. Wu, X. Li, and J. Chen, Opt. Express **27**, 32970-32983 (2019).
5. B. J. Shastri, A. N. Tait, T. Ferreira de Lima, W. H. P. Pernice, H. Bhaskaran, C. D. Wright, and P. R. Prucnal, Nature Photonics **15**, 102–114 (2021).
6. W. Bogaerts, D. Pérez, J. Capmany, D. A. B. Miller, J. Poon, D. Englund, F. Morichetti, and A. Melloni, Nature **586**, 207–216 (2020).
7. S. Chung, M. Nakai, and H. Hashemi, Opt. Express **27**, 13430-13459 (2019).
8. S. A. Miller, Y.-C. Chang, C. T. Phare, M. C. Shin, M. Zadka, S. P. Roberts, B. Stern, X. Ji, A. Mohanty, O. A. J. Gordillo, U. D. Dave, and M. Lipson, Optica **7**, 3-6 (2020).
9. R. Dekker, N. Usechak, M. Forst, and A. Driessen, J. Phys. D: Appl. Phys. **40**(14), R249–R271 (2007).
10. P. Sun and R. M. Reano, Opt. Express **18**, 8406-8411 (2010).
11. M. Jacques, A. Samani, E. El-Fiky, D. Patel, Z. Xing, and D. V. Plant, Opt. Express **27**, 10456-10471 (2019).
12. Q. Fang, J. F. Song, T.-Y. Liow, H. Cai, M. B. Yu, G. Q. Lo, and D.-L. Kwong, IEEE Photon. Technol. Lett. **23**(8), 525-527 (2011).
13. F. Duan, K. Chen, D. Chen, and Y. Yu, Opt. Lett. **46**, 234-237 (2021).
14. N. J. Martinez, C. T. DeRose, R. Jarecki, A. L. Starbuck, A. T. Pomerene, D. C. Trotter, and A. L. Lentine, *IEEE Optical Interconnects Conference* (*OI*, 2017), pp. 15–16.
15. R. Zhang, Y. He, Y. Zhang, S. An, Q. Zhu, X. Li and Y. Su, Nanophotonics **10**(2), 937-945 (2021).
16. F. A. Memon, F. Morichetti, and A. Melloni, ACS Photonics **5**(7), 2755-2759 (2018).
17. J.-M. Brosi, C. Koos, L. C. Andreani, M. Waldow, J. Leuthold, and W. Freude, Opt. Express **16**, 4177-4191 (2008).
18. L. Alloatti, R. Palmer, S. Diebold, K. P. Pahl, B. Chen, R. Dinu, M. Fournier, J.-M. Fedeli, T. Zwick, W. Freude, C. Koos, and J. Leuthold, Light: Sci. Appl. **3**, e173 (2014).
19. C. Kieninger, Y. Kutuvantavida, H. Miura, J. N. Kemal, H. Zwickel, F. Qiu, M. Lauermann, W. Freude, S. Randel, S. Yokoyama, and C. Koos, Opt. Express **26**, 27955-27964 (2018).
20. C. Kieninger, Y. Kutuvantavida, D. L. Elder, S. Wolf, H. Zwickel, M. Blaicher, J. N. Kemal, M. Lauermann, S. Randel, W. Freude, L. R. Dalton, and C. Koos, Optica **5**, 739-748 (2018).
21. Y. Xing, T. Ako, J. P. George, D. Korn, H. Yu, P. Verheyen, M. Pantouvaki, G. Lepage, P. Absil, A. Ruocco, C. Koos, J. Leuthold, Kristiaan Neyts, J. Beeckman, and W. Bogaerts, IEEE Photonics Technol. Lett. **27**(12), 1269-1272 (2015).
22. L.-Y. Chiang, C.-T. Wang, S. Pappert, and P. K.L. Yu, IEEE Photonics Technol. Lett. **33**(15), 796-799 (2021).
23. L. Van Iseghem, E. Picavet, A. Y. Takabayashi, P. Edinger, U. Khan, P. Verheyen, N. Quack, K. B. Gylfason, K. De Buysser, J. Beeckman, and W Bogaerts, Opt. Mater. Express **12**, 2181-2198 (2022).
24. L.-Y. Chiang, C.-T. Wang, T.-S. Lin, S. Pappert, and P. Yu, Opt. Express **28**, 29345-29356 (2020).
25. V. Tkachenko, A. Marino, and G. Abbate, Molecular Crystals and Liquid Crystals **527**, 80-91 (2010).
26. R. Ding, T. Baehr-Jones, W.-J. Kim, X. Xiong, R. Bojko, J.-M. Fedeli, M. Fournier, and M. Hochberg, Opt. Express **18**, 25061-25067 (2010).
27. M. R. Watts, J. Sun, C. DeRose, D. C. Trotter, R. W. Young, and G. N. Nielson, Opt. Lett. **38**(5), 733–735 (2013).
28. P. Dhara and R. Mukherjee, RSC Adv. **9**, 21685–21694 (2019).
29. Y. Atsumi, K. Watabe, N. Uda, N. Miura, and Y. Sakakibara, Opt. Express **27**, 8756-8767 (2019).
30. L.-Y. Chiang, H.-C. Liao, C.-T. Wang, S. Pappert, and P. K.L. Yu, *Photonics in Switching and Computing*, (*PSC*, 2021), Tu3A.3.
31. P. J. Bock, P. Cheben, J. H. Schmid, J. Lapointe, A. Delâge, S. Janz, G. C. Aers, D.-X. Xu, A. Densmore, and T. J. Hall, Opt. Express **18**, 20251-20262 (2010).
32. M. Asheghi, M. Touzelbaev, K. Goodson, Y. Leung, and S. Wong, J. Heat Trans. **120**(1), 30–36 (1998).
33. G. Ahlers, D. S. Cannell, L. I. Berge, and S. Sakurai, Physical Review E **49**(1), 545 (1994).
34. T. Kita and M. Mendez-Astudillo, J. Lightwave Technol. **39**, 5054-5060 (2021).
35. H. Qiu, Y. Liu, C. Luan, D. Kong, X. Guan, Y. Ding, and H. Hu, Opt. Lett. **45**, 4806-4809 (2020).
36. L.-Y. Chiang, C.-T. Wang, S. Pappert, and P. K.L. Yu, *5th IEEE Electron Devices Technology & Manufacturing Conference* (*EDTM*, 2021), WE2P2-3.

**References with titles**

1. Q. Cheng, M. Bahadori, M. Glick, S. Rumley, and K. Bergman, "Recent advances in optical technologies for data centers: a review," Optica **5**, 1354-1370 (2018).
2. B. G. Lee and N. Dupuis, "Silicon Photonic Switch Fabrics: Technology and Architecture," J. Light. Technol. **37**, 6-20 (2019).
3. C.-P. Hsu, B. Li, B. Solano-Rivas, A. R. Gohil, P. H. Chan, A. D. Moore, and V. Donzella, "A Review and Perspective on Optical Phased Array for Automotive LiDAR," IEEE J. Sel. Top. Quantum Electron. **27**(1), 1-16 (2021).
4. C. Li, X. Cao, K. Wu, X. Li, and J. Chen, "Lens-based integrated 2D beam-steering device with defocusing approach and broadband pulse operation for Lidar application," Opt. Express **27**, 32970-32983 (2019).
5. B. J. Shastri, A. N. Tait, T. Ferreira de Lima, W. H. P. Pernice, H. Bhaskaran, C. D. Wright, and P. R. Prucnal, "Photonics for artificial intelligence and neuromorphic computing," Nature Photonics **15**, 102–114 (2021).
6. W. Bogaerts, D. Pérez, J. Capmany, D. A. B. Miller, J. Poon, D. Englund, F. Morichetti, and A. Melloni, "Programmable photonic circuits," Nature **586**, 207–216 (2020).
7. S. Chung, M. Nakai, and H. Hashemi, "Low-power thermo-optic silicon modulator for large-scale photonic integrated systems," Opt. Express **27**, 13430-13459 (2019).
8. S. A. Miller, Y.-C. Chang, C. T. Phare, M. C. Shin, M. Zadka, S. P. Roberts, B. Stern, X. Ji, A. Mohanty, O. A. J. Gordillo, U. D. Dave, and M. Lipson, "Large-scale optical phased array using a low-power multi-pass silicon photonic platform," Optica **7**, 3-6 (2020).
9. R. Dekker, N. Usechak, M. Forst, and A. Driessen, "Ultrafast nonlinear all-optical processes in silicon-on-insulator waveguides," J. Phys. D: Appl. Phys. **40**(14), R249–R271 (2007).
10. P. Sun and R. M. Reano, "Submilliwatt thermo-optic switches using free-standing silicon-on-insulator strip waveguides," Opt. Express **18**, 8406-8411 (2010).
11. M. Jacques, A. Samani, E. El-Fiky, D. Patel, Z. Xing, and D. V. Plant, "Optimization of thermo-optic phase-shifter design and mitigation of thermal crosstalk on the SOI platform," Opt. Express **27**, 10456-10471 (2019).
12. Q. Fang, J. F. Song, T.-Y. Liow, H. Cai, M. B. Yu, G. Q. Lo, and D.-L. Kwong, "Ultralow Power Silicon Photonics Thermo-Optic Switch With Suspended Phase Arms," IEEE Photon. Technol. Lett. **23**(8), 525-527 (2011).
13. F. Duan, K. Chen, D. Chen, and Y. Yu, "Low-power and high-speed 2 × 2 thermo-optic MMI-MZI switch with suspended phase arms and heater-on-slab structure," Opt. Lett. **46**, 234-237 (2021).
14. N. J. Martinez, C. T. DeRose, R. Jarecki, A. L. Starbuck, A. T. Pomerene, D. C. Trotter, and A. L. Lentine, "Substrate removal for ultra efficient silicon heater-modulators," *IEEE Optical Interconnects Conference* (*OI*, 2017), pp. 15–16.
15. R. Zhang, Y. He, Y. Zhang, S. An, Q. Zhu, X. Li and Y. Su, "Ultracompact and low-power-consumption silicon thermo-optic switch for high-speed data," Nanophotonics **10**(2), 937-945 (2021).
16. F. A. Memon, F. Morichetti, and A. Melloni, "High Thermo-Optic Coefficient of Silicon Oxycarbide Photonic Waveguides," ACS Photonics **5**(7), 2755-2759 (2018).
17. J.-M. Brosi, C. Koos, L. C. Andreani, M. Waldow, J. Leuthold, and W. Freude, "High-speed low-voltage electro-optic modulator with a polymer-infiltrated silicon photonic crystal waveguide," Opt. Express **16**, 4177-4191 (2008).
18. L. Alloatti, R. Palmer, S. Diebold, K. P. Pahl, B. Chen, R. Dinu, M. Fournier, J.-M. Fedeli, T. Zwick, W. Freude, C. Koos, and J. Leuthold, "100 GHz silicon–organic hybrid modulator," Light: Sci. Appl. **3**, e173 (2014).
19. C. Kieninger, Y. Kutuvantavida, H. Miura, J. N. Kemal, H. Zwickel, F. Qiu, M. Lauermann, W. Freude, S. Randel, S. Yokoyama, and C. Koos, "Demonstration of long-term thermally stable silicon-organic hybrid modulators at 85 °C," Opt. Express **26**, 27955-27964 (2018).
20. C. Kieninger, Y. Kutuvantavida, D. L. Elder, S. Wolf, H. Zwickel, M. Blaicher, J. N. Kemal, M. Lauermann, S. Randel, W. Freude, L. R. Dalton, and C. Koos, "Ultra-high electro-optic activity demonstrated in a silicon-organic hybrid modulator," Optica **5**, 739-748 (2018).
21. Y. Xing, T. Ako, J. P. George, D. Korn, H. Yu, P. Verheyen, M. Pantouvaki, G. Lepage, P. Absil, A. Ruocco, C. Koos, J. Leuthold, Kristiaan Neyts, J. Beeckman, and W. Bogaerts, "Digitally Controlled Phase Shifter Using an SOI Slot Waveguide With Liquid Crystal Infiltration," IEEE Photonics Technol. Lett. **27**(12), 1269-1272 (2015).
22. L.-Y. Chiang, C.-T. Wang, S. Pappert, and P. K.L. Yu, "Ultralow-$V_\pi L$ Silicon Electro-Optic Directional Coupler Switch With a Liquid Crystal Cladding," IEEE Photonics Technol. Lett. **33**(15), 796-799 (2021).
23. L. Van Iseghem, E. Picavet, A. Y. Takabayashi, P. Edinger, U. Khan, P. Verheyen, N. Quack, K. B. Gylfason, K. De Buysser, J. Beeckman, and W Bogaerts, "Low power optical phase shifter using liquid crystal actuation on a silicon photonics platform," Opt. Mater. Express **12**, 2181-2198 (2022).
24. L.-Y. Chiang, C.-T. Wang, T.-S. Lin, S. Pappert, and P. Yu, "Highly sensitive silicon photonic temperature sensor based on liquid crystal filled slot waveguide directional coupler," Opt. Express **28**, 29345-29356 (2020).
25. V. Tkachenko, A. Marino, and G. Abbate, "Study of Nematic Liquid Crystals by Spectroscopic Ellipsometry," Molecular Crystals and Liquid Crystals **527**, 80-91 (2010).
26. R. Ding, T. Baehr-Jones, W.-J. Kim, X. Xiong, R. Bojko, J.-M. Fedeli, M. Fournier, and M. Hochberg, "Low-loss strip-loaded slot waveguides in Silicon-on-Insulator," Opt. Express **18**, 25061-25067 (2010).
27. M. R. Watts, J. Sun, C. DeRose, D. C. Trotter, R. W. Young, and G. N. Nielson, "Adiabatic thermo-optic Mach-Zehnder switch," Opt. Lett. **38**(5), 733–735 (2013).
28. P. Dhara and R. Mukherjee, "Phase transition and dewetting of a 5CB liquid crystal thin film on a topographically patterned substrate," RSC Adv. **9**, 21685–21694 (2019).
29. Y. Atsumi, K. Watabe, N. Uda, N. Miura, and Y. Sakakibara, "Initial alignment control technique using on-chip groove arrays for liquid crystal hybrid silicon optical phase shifters," Opt. Express **27**, 8756-8767 (2019).
30. L.-Y. Chiang, H.-C. Liao, C.-T. Wang, S. Pappert, and P. K.L. Yu, "Enhanced Thermo-Optic Efficiency of Silicon Photonic Switch with Passively Aligned Nematic Liquid Crystals," *Photonics in Switching and Computing*, (*PSC*, 2021), Tu3A.3.
31. P. J. Bock, P. Cheben, J. H. Schmid, J. Lapointe, A. Delâge, S. Janz, G. C. Aers, D.-X. Xu, A. Densmore, and T. J. Hall, "Subwavelength grating periodic structures in silicon-on-insulator: a new type of microphotonic waveguide," Opt. Express **18**, 20251-20262 (2010).
32. M. Asheghi, M. Touzelbaev, K. Goodson, Y. Leung, and S. Wong, "Temperature-dependent thermal conductivity of single-crystal silicon layers in SOI substrates," J. Heat Trans. **120**(1), 30–36 (1998).
33. G. Ahlers, D. S. Cannell, L. I. Berge, and S. Sakurai, "Thermal conductivity of the nematic liquid crystal 4-n-pentyl-4'-cyanobiphenyl," Physical Review E **49**(1), 545 (1994).
34. T. Kita and M. Mendez-Astudillo, "Ultrafast Silicon MZI Optical Switch With Periodic Electrodes and Integrated Heat Sink," J. Lightwave Technol. **39**, 5054-5060 (2021).
35. H. Qiu, Y. Liu, C. Luan, D. Kong, X. Guan, Y. Ding, and H. Hu, "Energy-efficient thermo-optic silicon phase shifter with well-balanced overall performance," Opt. Lett. **45**, 4806-4809 (2020).
36. L.-Y. Chiang, C.-T. Wang, S. Pappert, and P. K.L. Yu, "Efficient silicon photonic waveguide switches for chip-scale beam steering applications," *5th IEEE Electron Devices Technology & Manufacturing Conference* (*EDTM*, 2021), WE2P2-3.

# Silicon-organic hybrid thermo-optic switch based on a slot waveguide directional coupler: supplemental document

**Experimental setup**

Fig. S1 shows a schematic diagram of the experimental setup used to measure the proposed switch's optical transmission and transient responses. A temperature control chuck was used to hold the device under test (DUT) and maintain the background temperature at 25°C during the measurement. A tunable laser with 1550 nm wavelength was applied through a single-mode fiber (SMF), an Erbium-Doped Fiber Amplifier (EDFA), and a polarization controller (PC) to get TE-polarized input light. A lensed fiber was incorporated to couple the TE input light into the silicon waveguide through an edge coupler. To measure the optical transmission, a DC power supply was connected to the electrodes of the device through DC probes. An objective lens collimated the light transmitted through the device and exited the silicon waveguide outputs. The output light beam was then filtered by an iris and a TE polarizer before being detected by a photodetector (PD). The PD was connected to a Data Acquisition (DAQ) system to record the results. A voltage sweeping was applied from the DC power supply to obtain the transmission curves as a function of electrical power.

To measure the transient responses, the DC power supply and DAQ were replaced by a function generator and an oscilloscope, respectively. Square-wave signals from a function generator were also applied to the device through probing. The transient responses were recorded by the oscilloscope for extraction of the time constants.

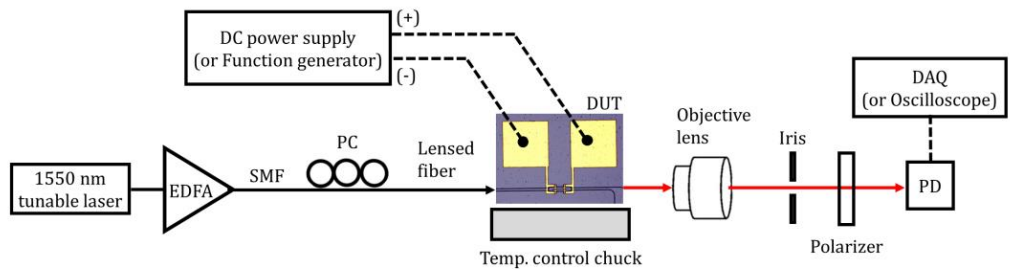

Fig. S1. Schematic diagram of the experimental setup